\documentclass{article}
\usepackage{ijcai22}

\usepackage{natbib}
\usepackage{times}
\usepackage{soul}
\usepackage{url}
\usepackage[hidelinks]{hyperref}
\usepackage[utf8]{inputenc}
\usepackage[small]{caption}
\usepackage{graphicx}
\usepackage{amsmath}
\usepackage{amsthm}
\usepackage{booktabs}
\usepackage{algorithm}
\usepackage{algorithmic}
\usepackage{xcolor}
\usepackage{balance}
\usepackage{amsfonts,amssymb}
\usepackage{makecell}
  \newcommand\figcaption{\def\@captype{figure}\caption}
  \newcommand\tabcaption{\def\@captype{table}\caption}
\makeatother

\title{Adversarial Bi-Regressor Network for Domain Adaptive Regression}

\author{
Haifeng Xia$^\dag$\thanks{The work was done during the internship at MERL.} \and
Pu Wang$^\sharp$ \and
Toshiaki Koike-Akino$^\sharp$ \and
Ye Wang$^\sharp$ \and
Philip Orlik$^\sharp$ \and
Zhengming Ding$^\dag$
\affiliations
$^\dag$Department of Computer Science, Tulane University, New Orleans, LA\\
$^\sharp$Mitsubishi Electric Research Laboratories (MERL), Cambridge, MA
\emails
\{hxia, zding1\}@tulane.edu,
\{pwang, koike, yewang, porlik\}@merl.com
}

\begin{document}

\maketitle

\begin{abstract}
Domain adaptation (DA) aims to transfer the knowledge of a well-labeled source domain to facilitate unlabeled target learning. When turning to specific tasks such as indoor (Wi-Fi) localization, it is essential to learn a cross-domain regressor to mitigate the domain shift. This paper proposes a novel method Adversarial Bi-Regressor Network (\textbf{ABRNet}) to seek more effective cross-domain regression model. Specifically, a discrepant bi-regressor architecture is developed to maximize the difference of bi-regressor to discover uncertain target instances far from the source distribution, and then an adversarial training mechanism is adopted between feature extractor and dual regressors to produce domain-invariant representations. To further bridge the large domain gap, a domain-specific augmentation module is designed to synthesize two source-similar and target-similar intermediate domains to gradually eliminate the original domain mismatch. The empirical studies on two cross-domain regressive benchmarks illustrate the power of our method on solving the domain adaptive regression (DAR) problem.
\end{abstract}

\section{Introduction}
Deep neural network has become popular on solving many practical scenarios whether the specific task is classification or regression \cite{sellami2022deep,xu2014regression}. However, considerable network parameters to be optimized means we have to collect extensive training samples with precise annotation, which tends to be time-consuming and laborious for real applications \cite{krizhevsky2012imagenet}. The intuition is to utilize the off-the-shelf label-sufficient data to accumulate knowledge and directly adapt it to solve other specific and relevant challenges \cite{awais2021adversarial,xia2021semi}. But the knowledge transfer is negatively obstructed by the distribution discrepancy across training and test sets so that the well-learned network suffers from performance degradation \cite{jing2021towards,xia2021adaptive}. The dilemma motivates the development of unsupervised domain adaptation (UDA) to bring knowledge from source domain to predict unlabeled target instances by learning domain-invariant representations. \cite{pan2019transferrable}.

\begin{figure}[t]
     \begin{center}
     \hspace{-0mm}\includegraphics[width=0.45\textwidth]{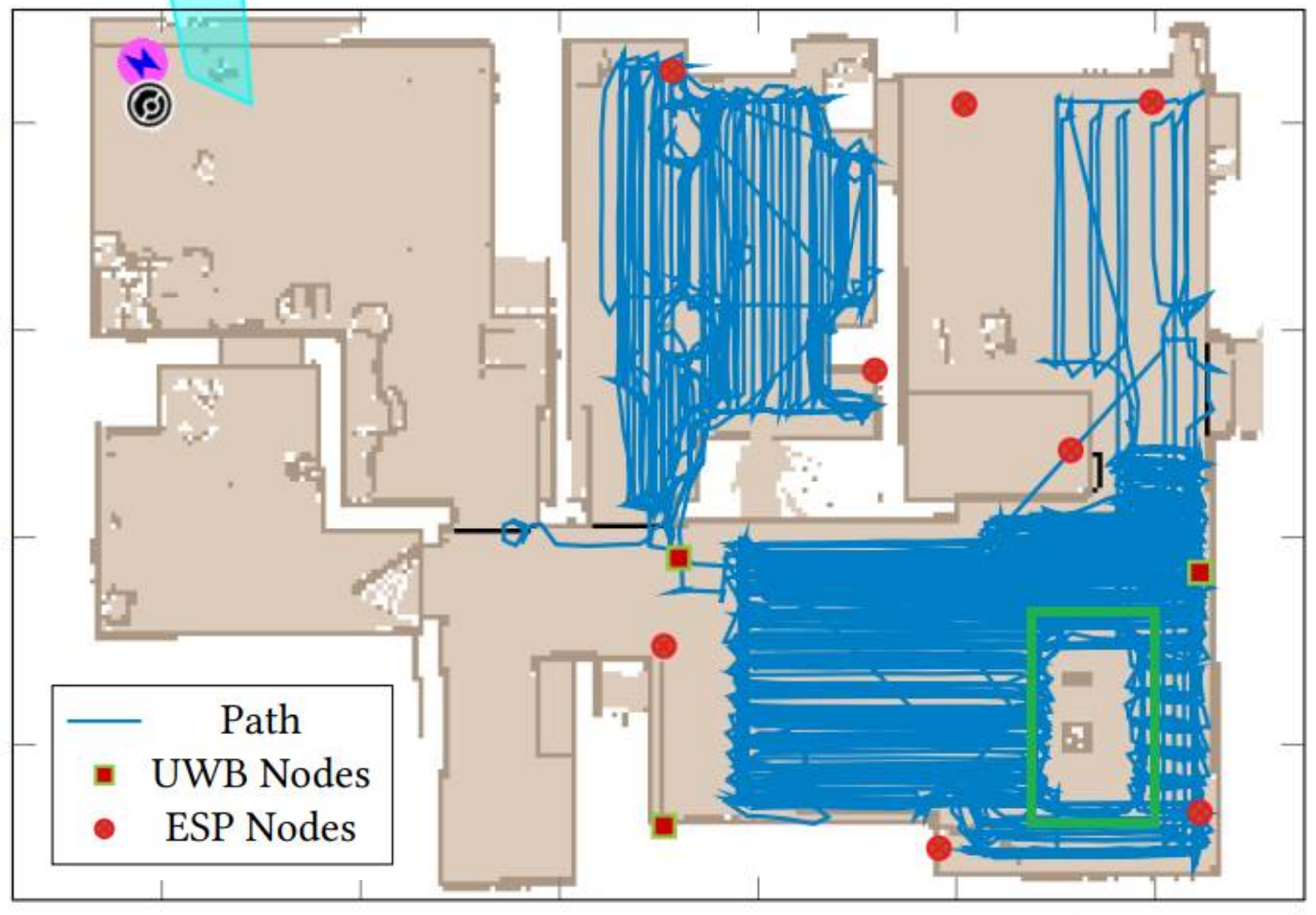}
        \end{center}
    \vspace{-0.1in}
    \caption{The floor map of the \emph{SPAWC2021 multi-modal indoor localization dataset} \cite{ArnoldSchaich21}, where red squares and red circles denote, respectively,  UWB and Wi-Fi anchors, while blue line denotes the path of a robot. The green box denotes the area where certain furniture was moved between offline fingerprinting and online test data collection.}
   \label{fig:floorplan}
  \vspace{-0.1in}
\end{figure}

\begin{figure*}[t]
    \centering
    \includegraphics[width=0.97\textwidth]{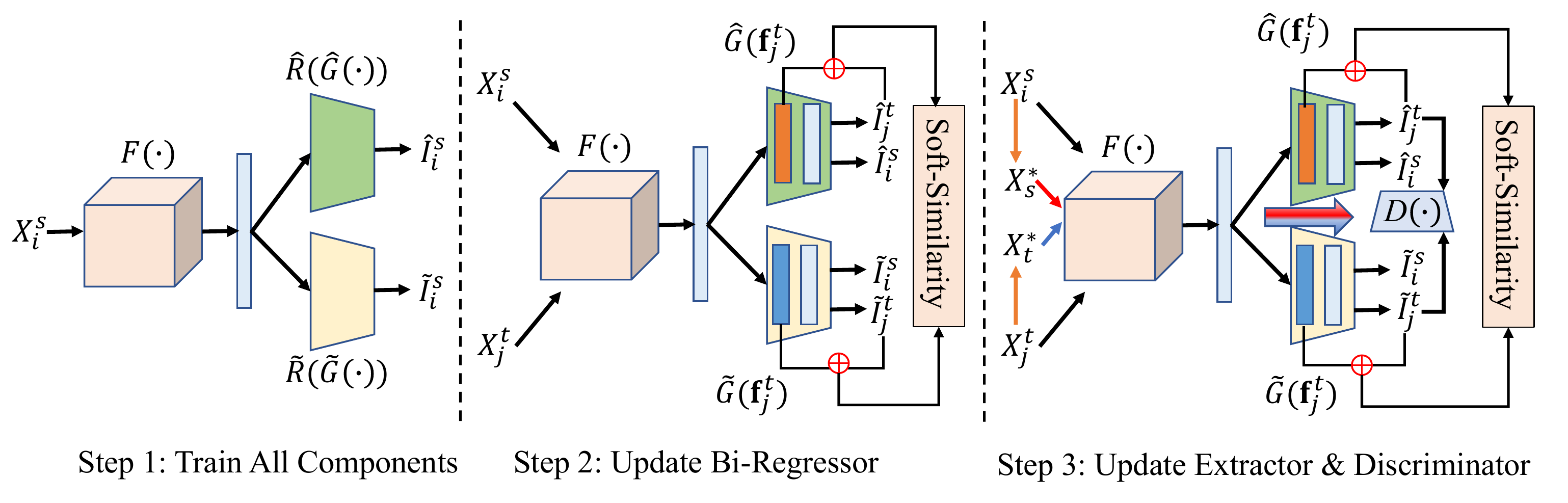}\vspace{-4mm}
    \caption{Overview of our proposed adversarial bi-regressors network (ABRNet) including feature generator $F(\cdot)$, two regressors $\{{\hat{G},\hat{R}}\}$, $\{{\widetilde{G},\widetilde{R}}\}$ and discriminator $D(\cdot)$. The main training process involves three stages. During the first step, ABRNet utilizes all source supervisions to train the network. For the second step, we introduce soft-similarity to maximize the bi-regressor difference. The third one aims to generate domain-invariant representations by updating feature generator and discriminator with the frozen regressors.}\vspace{-4mm}
    \label{framework}
\end{figure*}

Along with this direction, abundant UDA solutions have achieved promising performance in classification task \cite{na2021fixbi,chang2019domain,xia2020structure}. To the best of our knowledge, there are a few studies exploring the domain adaptation regression while they focus on the theoretical analysis in shallow regime without the benefit of deep neural architecture \cite{pan2010domain,mansour2009domain,yamada2014domain}. The regression task is of great importance and common in our reality, e.g., indoor wireless localization \cite{tong2021wi,zhou2021integrated,ArnoldSchaich21}. Concretely, a robot moves in a multi-room environment as shown in Figure \ref{fig:floorplan} and communicates with multi-modality anchors, e.g., Wi-Fi and ultra-wideband (UWB), placed at fixed locations denoted as red points. By processing the multi-modal radio frequency (RF) signals, one needs to predict the coordinates of robot in the floor map and make subsequent decisions for path planning and obstacle avoidance. Potential applications include smart home and autonomous factory where asset tracking is crucial. 

The current DAR works mainly focus on the visual regression problem, where convolutional network typically detects the boundary information of object beneficial for the domain adaptation \cite{lecun1998gradient,krizhevsky2012imagenet}. Thus, the relatively simple alignment strategies as DANN \cite{ganin2015unsupervised} easily learn the domain-invariant information for regressive task. And RSD \cite{pmlr-v139-chen21u} empirically finds the further control of feature scale learned from image data achieves better distribution alignment. However, the multi-modal RF signals are sensitive to environmental changes due to the fact that a subtle change in the environment may impact the RF signal propagation and result in signal variations at these RF anchors. This leads to a significant distribution shift across different scenarios and affecting the model generalization. And the relationship between multi-sensor RF signals and localization output is too implicit to be detected, which increases the difficulty of learning intrinsic representations across two domains. We empirically evaluate the existing UDA methods \cite{courty2017joint,li2020enhanced}, and notice they hardly improve the model generalization when solving indoor localization based on multi-sensor RF signals.

To overcome the practical challenges, this paper proposes a robust and effective \textit{Adversarial Bi-Regressor Network (ABRNet)}. Specifically, the ABRNet network consists of one feature generator and two same-architecture regressors. To identify the certain and uncertain target samples, we adopt the adversarial training strategy by maximizing the bi-regressor discrepancy to seek domain-invariant representations. However, it is difficult to directly align source and target domains in the presence of a large discrepancy. An intuitive solution is to synthesize samples in intermediate domains to bridge the original two domains and gradually eliminate their domain mismatch. The main contributions of our work are summarized three folds:
\begin{itemize}
    \item First, we propose an adversarial bi-regressor architecture to achieve a direct alignment across the source and target domains. Specially, the disagreement of dual regressors is maximized to discover target samples outside the source support and then learn domain-invariant features.
    \item Second, to fight off a situation where the cross-domain distribution discrepancy becomes relatively significant, as in the case of multi-modal indoor localization, we attempt to construct two intermediate domains between the source and target domains and gradually eliminate their mismatch to achieve distribution alignment.
    \item Finally, the experimental performance on multi-model RF signal and image benchmarks illustrates the effectiveness of our method on solving challenging DAR in real applications. Moreover, the empirical studies analyze the contributions of our designed modules.
\end{itemize}

\section{The Proposed Method}

\subsection{Problem Statement and Motivation}

Domain adaptive regression (DAR) aims to learn transferable knowledge from the label-sufficient source domain $\mathcal{D}_{s}=\{{\mathbf{X}_{i}^{s}, \mathbf{l}_{i}^{s}}\}_{i=1}^{n_{s}}$ to predict the continuous output for the target samples $\mathcal{D}_{t}=\{{\mathbf{X}_{i}^{t}}\}_{i=1}^{n_{t}}$, where $\mathbf{X}_{i}^{s/t}$, $\mathbf{l}_{i}^{s}$ corresponds to the instances and \textbf{annotations sampled from the continuous space}, and $n_{s/t}$ is the number of instances for each domain. Moreover, $\mathbf{X}_{i}^{s}$ and $\mathbf{X}_{i}^{t}$ are collected from the same sensors but with varying environment settings such as the change of furniture location, but share the identical continuous label space, e.g., 2D coordinate within the same floor map. 

Take the fingerprinting-based indoor localization as an example \cite{HeChan16,WangGao16,WangGao17,KoikeWang20,YuWang21}. During the offline fingerprinting phase, a robot moves in a pre-defined area, e.g., a multi-room apartment in Figure~\ref{fig:floorplan}, and registers  locations by collecting multi-modal RF signals from a set of fixed (Wi-Fi and UWB) anchors. Denoting the fingerprinted RF signals as $\mathbf{X}_{i}^{s}\in \mathbb{R}^{m\times d}$ where $m$ is the number of continuously fingerprinted samples during the offline fingerprinting phase, and $d$ is the dimension of the concatenation of multi-modal RF signals. Corresponding 2D coordinates $\mathbf{l}_{i}=(\mathbf{l}^{x}_{i}, \mathbf{l}^{y}_{i})$ of $m$ fingerprinted samples can be obtained from the robot using more costly on-board sensors such as Lidar and high-precision wheel encoder, where $\mathbf{l}^{x}_{i}\in [0, L], \mathbf{l}^{y}_{i}\in [0, W]$ and $L$, $W$ are the length and width of the floor respectively. Then, during the online test phase, new data $\mathbf{X}_{i}^{t}$ are collected under, possibly, changed environments such as the movement of furniture and the close/open of a door. As a result, well-trained localization models using the offline fingerprinted (source domain) data $\mathbf{X}_{i}^{s}$ may suffer from performance degradation when evaluating it on the test (target domain) samples $\mathbf{X}_{i}^{t}$ from changed environments. The problem of interest is to predict the continuous 2D coordinates of these new test data $\mathbf{X}_{i}^{t}$ by learning intrinsic and domain-invariant features from the offline signals.

Motivated by this consideration, this paper presents a novel method named Adversarial Bi-Regressors Network (ABRNet) to achieve distribution alignment for DAR setting. As shown in Figure \ref{framework}, ABRNet includes one feature extractor, two same-architecture regressors and one discriminator. The dual-regressor design allows the model to detect target samples outside source distribution and serves as one implicit discriminator to generate domain-invariant features. Moreover, to overcome the considerable distribution shift across source and target domains, ABRNet constructs source-similar and target-similar domains via the linear combination of original inputs and align them with the explicit discriminator to gradually eliminate the original cross-domain discrepancy.

\subsection{Bi-Regressor Discrepancy Discovery}

\begin{figure}
    \centering
    \includegraphics[width=0.49\textwidth]{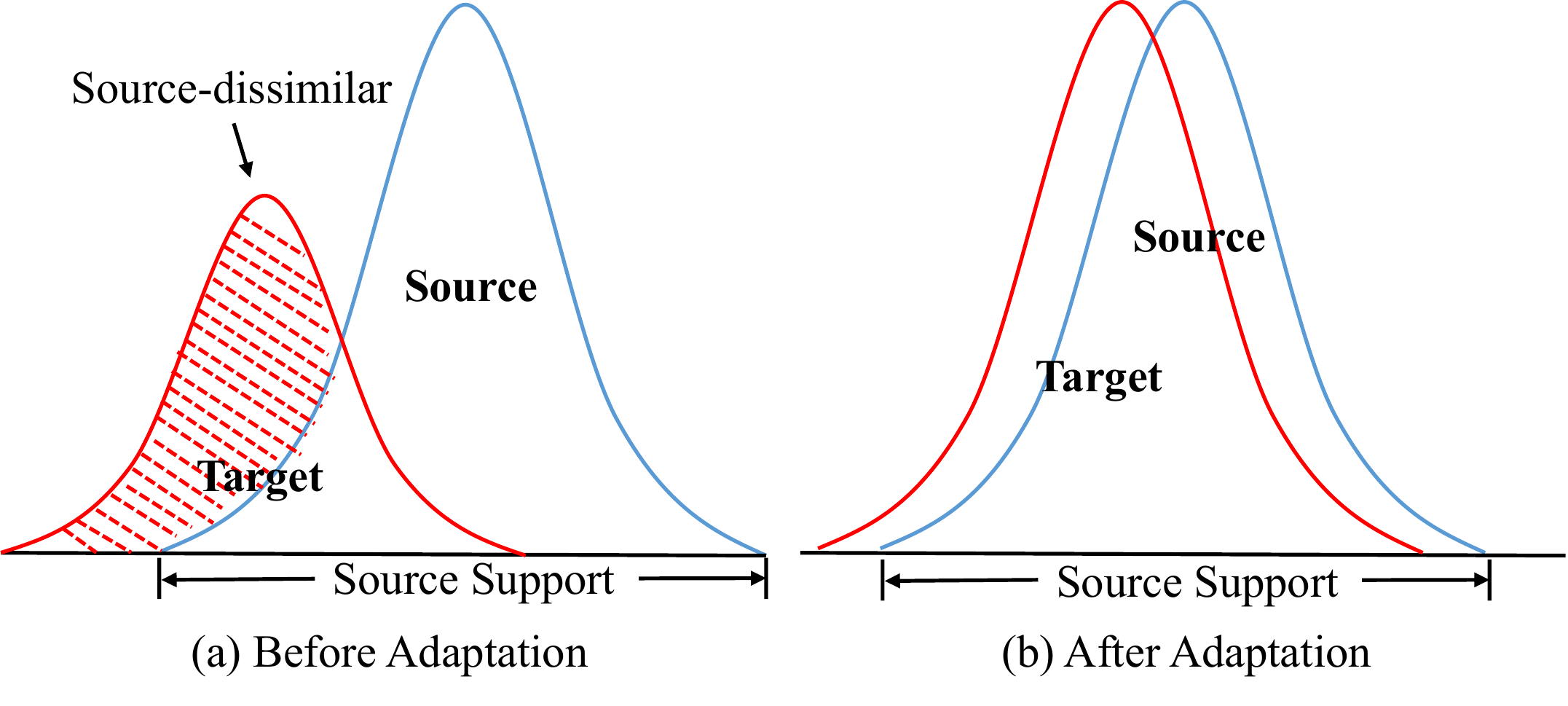}\vspace{-4mm}
    \caption{Before adaption (a): target samples far from the source distribution are named as source-dissimilar ones; After adaption (b):  target features are more aligned over the source support.}\vspace{-3mm}
    \label{adaption}
\end{figure}

The feature generator takes source and target samples as input to yield high-level feature representations, i.e., $\mathbf{f}^{s/t}_{i}=F(\mathbf{X}^{s/t}_{i}$). For specific tasks, we can consider the CNN or LSTM network architecture as the feature extractor. However, the significant cross-domain difference results in the feature distribution mismatch. For instance, in the indoor RF localization task, source and target features corresponding to the same location may become significantly different due to environment variations nearby. Thus, the regressor learned from the source supervision fails to accurately predict the coordinate of several target instances which are shown as shaded area in Figure \ref{adaption} and far from the source support. To achieve domain alignment, we first detect these target samples and adjust feature extractor to generate them within the source support. For the convenience of illustration, we name these target samples as source-dissimilar ones.

According to the voting strategy, two regressors easily produce inconsistent prediction on source-dissimilar target instances. Thus, we utilize the prediction disagreement to detect those target samples. Specifically, the introduced two regressors are equipped with the same network architecture, i.e., two fully-connected layers transforming the high-level feature into the final prediction via $\hat{\mathbf{g}}_{i}^{s/t}=\hat{G}(\mathbf{f}_{i}^{s/t})$ or $\widetilde{\mathbf{g}}_{i}^{s/t}=\widetilde{G}(\mathbf{f}_{i}^{s/t})$ and $\hat{\mathbf{l}}_{i}^{s/t}=\hat{R}(\hat{\mathbf{g}}_{i}^{s/t})$ or $\widetilde{\mathbf{l}}_{i}^{s/t}=\widetilde{R}(\widetilde{\mathbf{g}}_{i}^{s/t})$, where $\{{\hat{G}, \hat{R}}\}$ and $\{{\widetilde{G}, \widetilde{R}}\}$ are the components of two regressors, respectively. If the source and target regressors provide different predictions for the same target samples, we consider them as source-dissimilar ones with high confidence. The prediction discrepancy means that these two regressors involve different predictive abilities for target instances. For this objective, we need to maximize the discrepancy of two regressor. Different from the classification task, the final output of the regressors fail to reflect the probability distribution of samples in the continuous label space. Although the maximization of the final outputs of two regressors makes them different, such an operation easily misleads the regressor to deliberately search for solutions that are far from the ground-truth, leading to degraded performance of the regressor.

On the other hand, it is observed that the first output of regressor $\mathbf{g}_{i}^{t}$ ($\mathbf{g}\in \{{\hat{\mathbf{g}}, \widetilde{\mathbf{g}}}\}$) indicates the distribution of hidden feature space. Thus, we consider $\mathbf{g}_{i}^{t}$ as the prior condition for the final output and concatenate them as $\hat{\mathbf{h}}_{i}^{t}=[\hat{\mathbf{g}}_{i}^{t}, \hat{\mathbf{l}}_{i}^{t}]$ or $\widetilde{\mathbf{h}}_{i}^{t}=[\widetilde{\mathbf{g}}_{i}^{t}, \widetilde{\mathbf{l}}_{i}^{t}]$. To quantify the similarity between $\hat{\mathbf{h}}_{i}^{t}$ and $\widetilde{\mathbf{h}}_{i}^{t}$ that reflects the bi-regressor discrepancy, one option is to use the Jaccard Similarity Coefficient (IoU score) from the field of object detection using visual sensors \cite{gupta2014learning}. However, the non-differentiable nature of the IoU score leads to the optimization of the network weights problematic. Alternatively, a differentiable soft-similarity is developed here for this purpose:
\begin{equation}
    \label{IoU_variant}
    \mathcal{L}_{s} = \sum_{i=1}^{n_{t}} \frac{\langle \hat{\mathbf{h}}_{i}^{t}, \widetilde{\mathbf{h}}_{i}^{t}\rangle}{\sum_{d}(\hat{\mathbf{h}}_{i}^{t}+\widetilde{\mathbf{h}}_{i}^{t}-\hat{\mathbf{h}}_{i}^{t}\otimes\widetilde{\mathbf{h}}_{i}^{t})},
\end{equation}
where $\langle \cdot, \cdot \rangle$ is the inner-product suggesting the intersection of $\hat{\mathbf{h}}_{i}^{t}$ and $\widetilde{\mathbf{h}}_{i}^{t}$, $\otimes$ is the element-wise product, and $d$ is the dimension size of $\mathbf{h}_{i}^{t}$ ($\mathbf{h} \in\{{\hat{\mathbf{h}}, \widetilde{\mathbf{h}}}\}$). Note that $\mathbf{g}_{i}^{t}$ is activated with the Sigmoid function before computing this metric. The minimization of Eq. \eqref{IoU_variant} decreases the intersection between $\hat{\mathbf{g}}_{i}^{t}$ and $\widetilde{\mathbf{g}}_{i}^{t}$ to further increase the difference of dual regressors, and vice verse. With Eq. (\ref{IoU_variant}), we can achieve the domain alignment with the bi-regressor in the following three steps.

First, due to the accessibility of source labels, it is simple to train the entire network to learn and adapt the source support set via a supervised manner as:
\begin{equation}
    \label{step1}
    \min_{F,\hat{G}, \widetilde{G}, \hat{R}, \widetilde{R}} \mathcal{L}_{r} = \sum_{i=1}^{n_{s}}\Big( \mathcal{L}_{r}^{*}(\hat{\mathbf{l}}_{i}^{s}, \mathbf{l}_{i}^{s}) + \mathcal{L}_{r}^{*}(\widetilde{\mathbf{l}}_{i}^{s}, \mathbf{l}_{i}^{s})\Big),
\end{equation}
where $\mathcal{L}_{r}^{*}$ is the mean-square error (MSE) loss. This training manner likely yields two identical regressors. To avoid this issue, the second step is to maximize the difference of the bi-regressor with the aforementioned soft-similarity score to detect the source-dissimilar instances. Meanwhile, we still need to guarantee the predictive ability of model on the source samples. These considerations lead to the following objective function formulated as $\min_{\hat{G}, \widetilde{G}, \hat{R}, \widetilde{R}} \mathcal{L}_{r} + \mathcal{L}_{s}$. It is worth noting that the second step only updates the bi-regressor with the feature generator fixed. To this end, we can explore the bi-regressor discrepancy to discover the target samples outside the source support. The third step optimizes the feature extractor to move these target samples into the source support set by improving the overlap between $\hat{\mathbf{h}}^{t}_{i}$ and $\widetilde{\mathbf{h}}^{t}_{i}$ with the formulation as $\min_{F} -\mathcal{L}_{s}$.

\subsection{Intermediate Bi-Domain Alignment}

According to the above illustration, the conditional bi-regressor discrepancy maximizes the difference of two regressors to encourage the feature generator to produce domain-invariant representations. Thus, the essence of this module considers the combination of the two regressors as one discriminator to directly distinguish the source features from the target ones. However, the aforementioned training manner means these two regressors should be very similar with smaller difference which makes it invalid for considerable cross-domain shift. 

An intuitive solution is to find two intermediate domains between the source and target ones and gradually mitigate the shift of two new introduced domains to align original distributions. To this end, we further explore data augmentation to achieve this motivation. First, we adopt a fixed ratio $\lambda$ to linearly combine the source and target samples to synthesize source-similar and target-similar instances as:
\begin{equation}
    \label{combination}
    \mathbf{X}_{s}^{*} = \lambda \mathbf{X}_{i}^{s} + (1-\lambda) \mathbf{X}_{j}^{t},~~~
    \mathbf{X}_{t}^{*} = (1-\lambda) \mathbf{X}_{i}^{s} + \lambda \mathbf{X}_{j}^{t},
\end{equation}
where $\lambda$ is set to $0.7$ in all experiments. The combined instances gradually form two intermediate domains where one is closer to the source domain and the other is similar to the target domain. Second, we introduce an additional discriminator following the feature generator and utilize their adversarial relationship to mitigate the domain shift between source-similar and target-similar ones. Concretely, the augmented instances flow into the feature generator and two regressors with the corresponding output, i.e., $\mathbf{f}^{s/t}_{*}=F(\mathbf{X}_{s/t}^{*})$, $\hat{\mathbf{l}}_{*}^{s}=\hat{R}(\hat{G}(\widetilde{\mathbf{f}}^{s}_{*}))$ and $\widetilde{\mathbf{l}}_{*}^{t}=\widetilde{R}(\widetilde{G}(\widetilde{\mathbf{f}}_{*}^{t}))$. And then we adopt the adversarial loss as the following:
\begin{equation*}
    \label{adversarial_loss}
    \mathcal{L}_{adv} = \mathbb{E}_{X_{s}^{*}\sim\mathcal{D}_{s}^{*}} \log[D(\mathbf{f}^{s}_{*}, \hat{\mathbf{l}}_{*}^{s})]+\mathbb{E}_{X_{t}^{*}\sim \mathcal{D}_{t}^{*}} \log[1-D(\mathbf{f}^{t}_{*}, \widetilde{\mathbf{l}}_{*}^{t})],
\end{equation*}
where $\mathcal{D}_{s/t}^{*}$ denotes the source-similar or target-similar domain and $D(\cdot)$ is the domain discriminator. Since the augmented samples are derived from the original source and target samples, their adversarial training would trigger the domain alignment of the original two domains. Thus, our augmented domain alignment can compensate the previous bi-regressor alignment. 

\subsection{Overall Objectives}

So far, we have illustrated the details of our proposed ABRNet to seek domain-invariant feature representation for cross-domain regression. The overall optimization iterates the following three steps until the convergence is achieved or the maximum number of iterations is reached:
\begin{equation*}
    \left \{
    \begin{array}{lr}
    \textbf{Step 1}:  \min\limits_{F,\hat{G}, \widetilde{G}, \hat{R}, \widetilde{R}} \mathcal{L}_{r} = \sum\nolimits_{i=1}^{n_{s}} \mathcal{L}_{r}^{*}(\hat{\mathbf{l}}_{i}^{s}, \mathbf{l}_{i}^{s}) + \mathcal{L}_{r}^{*}(\widetilde{\mathbf{l}}_{i}^{s}, \mathbf{l}_{i}^{s}), \\
    \textbf{Step 2:} \quad \min\limits_{\hat{G}, \widetilde{G}, \hat{R}, \widetilde{R}} \mathcal{L}_{r} + \mathcal{L}_{s}, &\\
    \textbf{Step 3:} \quad \quad  \min\limits_{F} \mathcal{L}_{adv}-\mathcal{L}_{s}, \quad \max\limits_{D} \mathcal{L}_{adv}.
    \end{array}
    \right.
\end{equation*}

\section{Experiments}

\subsection{Experimental Steup}
\noindent \textbf{Datasets.} \textbf{1)} SPAWC2021 \cite{ArnoldSchaich21}  is a large-scale indoor localization dataset, where a robot moves in a multi-room apartment and uses the received multi-modal RF signals from the deployed devices to predict its coordinates in this floor plan. For each location, the recorded signals for this robot consists of RSSI (\textbf{R}, scalar), CSI (\textbf{C}, 64 values over  $64$ sub-carrier frequencies) from 11 Wi-Fi anchors, UWB (\textbf{U}) from 3 anchors recording range and power readings, and a 9-dimensional IMU (\textbf{I}) signal. SPAWC2021 consists a well-annotated Dataset1 (source domain) with 750$k$ samples and one unlabeled Dataset2 (target domain) with 650$k$ instances. Their differences are the movement of furniture around the green box in Figure \ref{fig:floorplan} and the change of data collection time, which results in significant distribution difference. \textbf{2)} dSprites \cite{higgins2016beta} is popular 2D synthetic image dataset suitable to unsupervised regressive domain adaptation and includes three domains as Color (\textbf{C}), Noisy (\textbf{N}) and Scream (\textbf{S}). And each domain involves 737,280 images. The regressive tasks in dSprites are to predict the scale and (x,y) coordinates.

\begin{table*}[t]
  \mbox{}\hfill
  \linespread{1.2}
  \begin{minipage}[t]{0.48\linewidth}
\caption{Mean square error (MSE) computed on SPAWC2021 dataset with the combination of two sensors under Unsupervised Regressive Domain Adaptation. The best performance is highlighted with \textbf{bold} type, while the second one is emphasized with \underline{underline}.} \vspace{-2mm}\small
\label{table:two_sensors}
\setlength{\tabcolsep}{6pt} 
\renewcommand{\arraystretch}{0.95} 
\begin{tabular}{ccccccccccc}
  \Xhline{1pt}
Method &RC &RU &RI &CU &CI &UI &Avg.\\ \hline
LSTM	&1.39	&1.08	&1.12	&1.39	&0.98	&2.43	&1.40\\
TCA	 &1.36	&0.97	&0.93	&1.32	&0.98	&2.22	&1.30\\
DAN	 &1.39	&1.02	&1.03	&1.34	&0.98	&2.29	&1.34\\
DANN 	&\underline{1.26}	&0.98	&\underline{0.92}	&1.26	&\underline{0.96}	&\underline{2.09}	&\underline{1.25}\\
JDOT 	&1.33	&0.97	&0.94	&1.28	&0.97	&2.13	&1.27\\
MCD	 &1.35	&\underline{0.96}	&0.98	&\underline{1.21}	&0.97	&2.10	&1.26\\
ETD	 &1.35	&0.98	&0.94	&1.23	&0.98	&2.11	&1.27\\ \hline
Ours	&\textbf{1.13}	&\textbf{0.84}	&\textbf{0.71}	&\textbf{0.98}	&\textbf{0.95}	&\textbf{1.58}	&\textbf{1.04}\\ 
  \Xhline{1pt}
\end{tabular}
  \end{minipage}\hfill
  \hspace{4mm}
  \begin{minipage}[t]{.48\linewidth}
\caption{MSE computed on SPAWC2021 dataset with the combination of three or four sensors under Unsupervised Regressive Domain Adaptation. The best performance and the second one is highlighted with \textbf{bold} type and \underline{underline} respectively.} \vspace{-2mm}\small
\label{table:multiple_sensors}
\setlength{\tabcolsep}{8pt} 
\renewcommand{\arraystretch}{0.95} 
\begin{tabular}{cccccccccc}
  \Xhline{1pt}
Method &RCU &RCI &RUI &CUI &RCUI &Avg.\\ \hline
LSTM	&1.04	&0.96	&0.89	&0.92	&0.88	&0.94\\
TCA	&1.02	&0.95	&0.87	&0.92	&0.79	&0.91\\
DAN	&\underline{0.96}	&0.95	&0.82	&0.86	&0.75	&0.87\\
DANN	&0.97	&{0.90}	&0.81	&0.87 &0.71	&\underline{0.85}\\
JDOT	&0.99	&\underline{0.89}	&0.88	&\underline{0.82}	&\underline{0.70}	&0.86\\
MCD	&1.01	&0.96	&\underline{0.78}	&0.85	&0.74	&0.87\\
ETD	&1.04	&0.94	&0.85	&0.86	&0.80	&0.90\\ \hline
Ours	&\textbf{0.85}	&\textbf{0.70}	&\textbf{0.57}	&\textbf{0.68}	&\textbf{0.54}	&\textbf{0.67}\\ 
  \Xhline{1pt}
\end{tabular}
  \end{minipage}\hfill
  \mbox{}\vspace{-1mm}
\end{table*}

\begin{table*}[!htbp]
\linespread{1.2} 
\centering\small
\caption{Mean absolute error (MAE) computed on dSprites dataset with six tasks under Unsupervised Regressive Domain Adaptation. The best performance is highlighted with \textbf{bold} type, while the second one is emphasized with \underline{underline}.} \vspace{-2mm}
\label{table:dsprites}
\setlength{\tabcolsep}{12pt} 
\renewcommand{\arraystretch}{1.0} 
\begin{tabular}{ccccccccccc}
  \Xhline{1pt}
Method &C$\to$N &C$\to$S &N$\to$C &N$\to$S &S$\to$C &S$\to$N &Avg.\\ \hline
Resnet-18	&0.94$\pm$0.06	&0.90$\pm$0.08	&0.16$\pm$0.02	&0.65$\pm$0.02	&0.08$\pm$0.01	&0.26$\pm$0.03	&0.498\\
TCA	        &0.94$\pm$0.03	&0.87$\pm$0.02	&0.19$\pm$0.02	&0.66$\pm$0.05	&0.10$\pm$0.02	&0.23$\pm$0.04	&0.498\\
DAN	        &0.70$\pm$0.05	&0.77$\pm$0.09	&\textbf{0.12$\pm$0.03}	&\textbf{0.50$\pm$0.05}	&\underline{0.06$\pm$0.02}	&0.11$\pm$0.04	&0.377\\
DANN	    &0.47$\pm$0.07	&0.46$\pm$0.07	&0.16$\pm$0.02	&0.65$\pm$0.05	&\textbf{0.05$\pm$0.00}	&0.10$\pm$0.01	&0.315\\
JDOT	    &0.86$\pm$0.03	&0.79$\pm$0.02	&0.19$\pm$0.02	&0.64$\pm$0.05	&0.10$\pm$0.02	&0.23$\pm$0.04	&0.468\\
MCD	        &0.81$\pm$0.09	&0.81$\pm$0.12	&0.17$\pm$0.12	&0.65$\pm$0.03	&0.07$\pm$0.02	&0.19$\pm$0.04	&0.450\\
AFN	        &1.00$\pm$0.04	&0.96$\pm$0.05	&0.16$\pm$0.03	&0.62$\pm$0.04	&0.08$\pm$0.01	&0.32$\pm$0.06	&0.523\\
RSD	        &\underline{0.32$\pm$0.02}	&\underline{0.35$\pm$0.02}	&0.16$\pm$0.02	&0.57$\pm$0.01	&0.08$\pm$0.01	&\underline{0.09$\pm$0.02}	&\underline{0.258}\\
\hline
Ours	&\textbf{0.20$\pm$0.02}	&\textbf{0.27$\pm$0.01}	&\underline{0.14$\pm$0.01}
&\underline{0.56$\pm$0.03}	&\textbf{0.05$\pm$0.01}	&\textbf{0.08$\pm$0.02}	&\textbf{0.217}\\ 
  \Xhline{1pt}
\end{tabular}
\vspace{-3mm}
\end{table*}

\noindent \textbf{Implementation details.} Due to the difference of data type, we adopt distinct backbones to learn feature with corresponding optimizer on two benchmarks. For \textbf{SPAWC2021}, when moving along with the trajectory, the current location of robot is related to the status of previous $t$ time steps. Considering this point, we adopt Bi-LSTM \cite{huang2015bidirectional} as feature generator to capture the temporal dependency in the high-level representations. The inputs of Bi-LSTM correspond to the signals of 10 continuous time steps and its output is a 512-dimensional vector for each instance. In addition, the varying environmental factors have different impacts on different sensors. To evaluate the effectiveness of our method with various signals, we consider the concatenation of the multi-modal signals as the input. It is worth noting that the dimensions of the processed RSSI, CSI, UWB and IMU are 8, 40, 6 and 9, respectively. The domain-specific regressors and the discriminator both comprise of two full-connection layers. In the implementation, the Adam optimizer \cite{kingma2014adam} with a learning rate of 0.001 is used to update all network components. For \textbf{dSprites}, we following the training strategy of RSD \cite{pmlr-v139-chen21u} utilize the pre-trained ResNet-18 as backbone to extract features and SGD with a momentum 0.95 to optimize the network architecture. The learning rate of the new added network modules is 10 times larger than that of the pre-trained module.

\noindent \textbf{Baselines.} To assess the power of our method on solving DAR, we compare it with the state-of-the-art domain adaptation methods which are appropriate for the regressive challenges. Specifically, the competitors are divided into two branches. The first one is metric-based solutions such as Transfer Component Analysis (\textbf{TCA}) \cite{pan2010domain}, Joint Distribution Optimal Transport (\textbf{JDOT}) \cite{courty2017joint}, Deep Adaptation Network (\textbf{DAN}) \cite{long2015learning}, Adaptive Feature Norm (\textbf{AFN}) \cite{xu2019larger} and Enhanced Transport Distance (\textbf{ETD}) \cite{li2020enhanced} and Representation Subspace Distance (\textbf{RSD})\footnote{Note that we attempt to evaluate \textbf{RSD} with the public code on SPAWC2021 dataset but the low-rank operation easily suffers from the vanishing gradient problem within a few iterations. Thus, we fail to report the results of RSD on SPAWC2021 dataset.} \cite{pmlr-v139-chen21u}. Another is domain confusion approaches such as Domain Adversarial Neural Network (\textbf{DANN}) \cite{ganin2015unsupervised} and Maximum Classifier Discrepancy (\textbf{MCD}) \cite{saito2018maximum}. The domain adaptation baselines are deployed in the identical network (LSTM or ResNet-18) and tested on both datasets.

\begin{figure*}[t]
    \centering
    \includegraphics[width=0.99\textwidth]{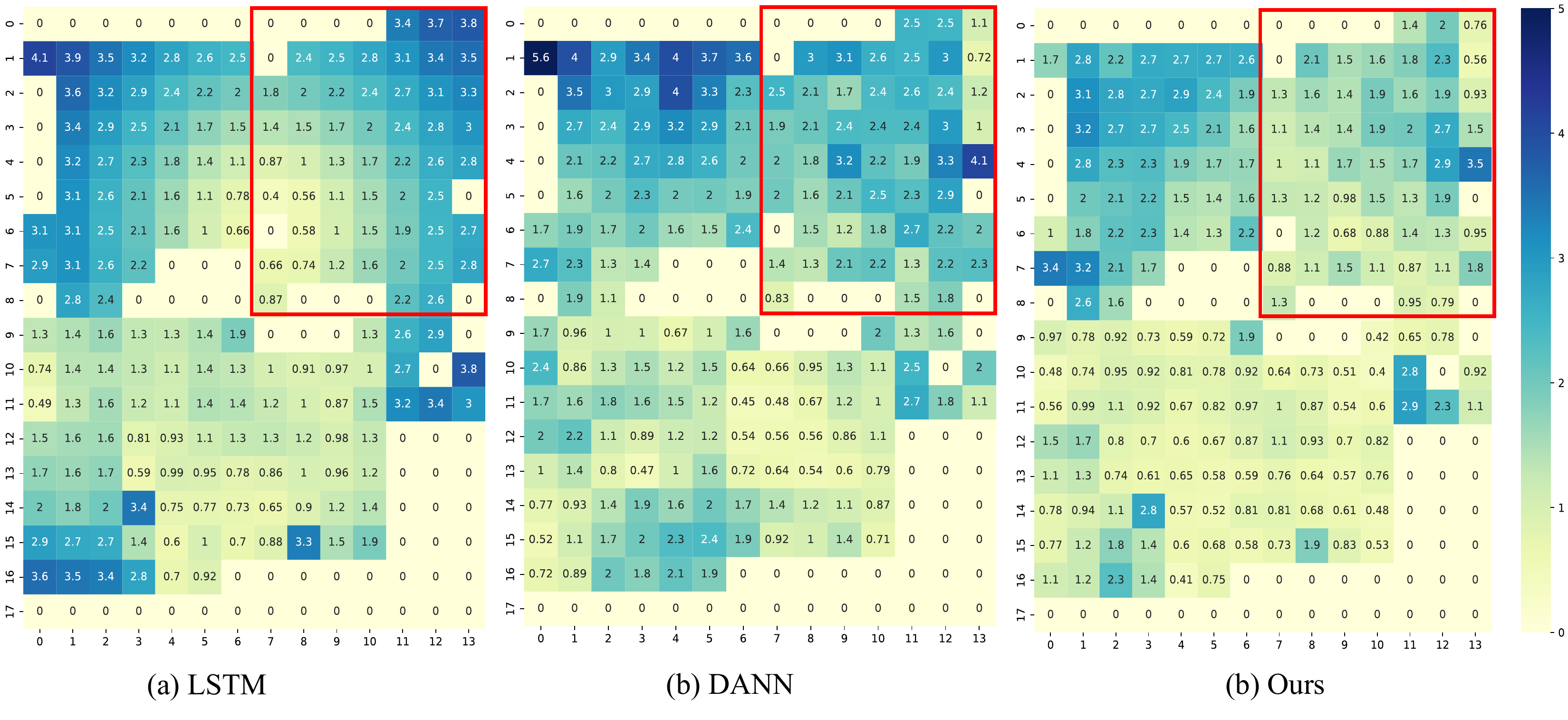} \vspace{-5mm}
    \caption{{Visualization of average MSEs over $50 \times 50$ $\text{cm}^2$ on the floor map. The $x$-axis and $y$-axis are divided several intervals with 0.5 meter, which forms the chessboard. The number of each grid is computed by taking the average MSE of all test samples falling into the grid.
    }}\vspace{-4mm}
    \label{vis}
\end{figure*}

\begin{figure*}[t]
    \centering
    \includegraphics[width=1\textwidth]{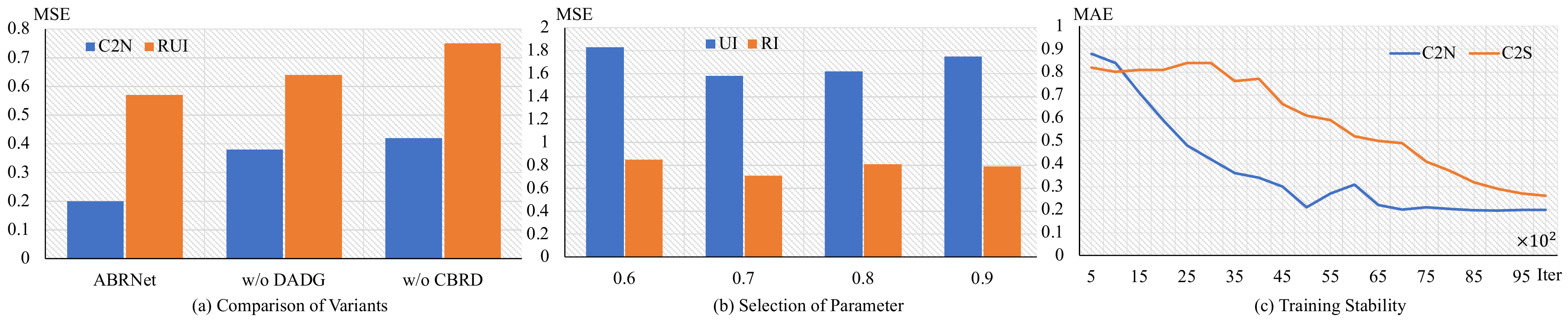} \vspace{-6mm}
    \caption{{Empirical Analysis: (a) Comparison of ABRNet variants, (b) Selection of parameter $\lambda$, and (c) Training stability with iterations.}} \vspace{-5mm}
    \label{ablation}
\end{figure*}

\subsection{Comparison Results}

Table \ref{table:two_sensors} and Table \ref{table:multiple_sensors} summarize the results of the mentioned methods on SPAWC2021 dataset with the concatenation of different signals. From both tables, several observations can be made as follows. \textbf{First}, our proposed method outperforms the baselines by a large margin on the average mean square error (MSE). Specially, when only two types of sensor are used, our strategy further improves the precision of localization prediction by comparing it with the second best result obtained by DANN (1.04 v.s. 1.25). It suggests that our method effectively eliminates the distribution discrepancy across Dataset1 and Dataset2 and transfers more source knowledge to target domain. Compared with DANN, our learning strategy explores the data augmentation to construct source-similar and target-similar samples and aligns them to achieve distribution matching. \textbf{Second}, it is simple to find that the model generalization is driven by the combination of different type of sensors. For example, when directly using the well-trained LSTM model into target domain, training model with CSI and IMU (CI) obtains better localization than that with UWB and IMU (UI) (0.98 v.s. 2.43). The main reason lies in that the signals UI are sensitive to varying external factors triggering considerable shift across source and target domains. Our proposed approach shows stronger mitigation of cross-domain discrepancy than the selected baselines. Especially, the comparison between our method and MCD with (1.58 v.s. 2.10) illustrates that our designed bi-regressor module better utilizes the maximization of discrepancy to discover the target samples far away from the source support and then minimize the dual-regressor difference to generate domain-invariant representations. \textbf{Third}, our method achieves the best localization precision (MSE:0.54) with the combination of all four type of sensors where the MSE of LSTM without any domain adaptations is about 0.88. The performance improvement is attributed to the efficient alignment of source and target distributions.

Table \ref{table:dsprites} reports the regressive performance on dSprites dataset. The distribution discrepancy in some tasks is smaller so that the well-learned source model (ResNet-18) shows promising prediction on target domain, e.g., N$\to$C and S$\to$C, where our method achieves comparable performance. With respect to more challenging tasks such as C$\to$S, the prediction ability of our model on target domain becomes much better than that of ResNet-18 with 0.27 v.s. 0.90. Moreover, compared with RSD, our model further reduces the MAE from 0.32 to 0.20 on task C$\to$N, which demonstrates that our model effectively eliminates the domain shift.

\subsection{Empirical Analysis}

\noindent \textbf{Visualization.} To better comprehend the performance gain of our method on SPAWC2021, Figure \ref{vis} specially visualizes the average MSE over 50$\times$50 $cm^{2}$ grids in the floor map and compare it with other baselines e.g., LSTM and DANN. The data description of SPAWC2021 shows for target domain, several cabinets were newly deployed in the area indicated by red box during the online data collection, leading to a distribution shift across source and target data. Thus, ResNet-18 and DANN give much worse localization performance than the proposed method for the target regressive task in the red box. Our method shows better capacity to distinguish source-dissimilar samples in the red box and significantly reduces the localization errors.

\noindent \textbf{Ablation Study \& Training Stability.} Our method includes two important operations: conditional bi-regressor discrepancy (CBRD) and data augmentation driven game (DADG). To analyse the contribution of each module, we first remove one part while keeping the other intact to form two variants: w/o DADG and w/o CBRD. From the results in Figure \ref{ablation}, it is seen that CBRD contributes more to the performance gain than DADG as the amount of localization error increase more w/o CBRD than w/o DADG. In addition, we analyse the performance sensitivity to different parameter ($\lambda$) selections on tasks UI and RI. From Figure \ref{ablation}, the model appears to be insensitive to $\lambda$ and achieves the best performance with $\lambda=0.7$. Finally, to assess the training stability, we record the MAE of target domain via target regressor over the number of iterations in Figure \ref{ablation} (c) which suggests the training process achieves convergence.

\section{Conclusion}
This paper explores domain adaptation regression (DAR) aiming to transfer source knowledge to provide continuous-label prediction for unlabeled target samples in real-world applications such as indoor localization. To this end, we propose a novel adversarial bi-regressor network (ABRNet) that consists of a feature generator, two regressors and a discriminator. First, we utilize the prediction disagreement of two regressors to identify target samples outside source distribution and then seek domain-invariant features depending on the adversarial relationship between feature generator and dual regressors. Moreover, considering the difficulty of direct alignment across two domains with large domain shift, ABRNet attempts to construct two intermediate domains with smaller discrepancy to gradually eliminate the original domain mismatch. Extensive experiments verify our method outperforms the state-of-the-art on solving DAR issue.

\balance
\newpage
\small
\bibliographystyle{named}
\bibliography{ABRNet}

\end{document}